\documentclass[artice,aps,pra,showpacs,twocolumn,superscriptaddress]{revtex4}
\usepackage{graphicx,color}
\usepackage{amsmath,amsthm,amsfonts,amssymb,bm}
\usepackage{times}
\usepackage{epsf}
\usepackage[colorlinks={true}]{hyperref}
\usepackage[T1]{fontenc}
\usepackage[utf8]{inputenc}

\hypersetup{citecolor={blue}, filecolor={blue}, linkcolor={blue}, urlcolor={blue}}

\newcommand{\commentold}[1]{}
\DeclareMathSymbol{:}{\mathpunct}{operators}{"3A}

\begin{document}

\title{Reducing the entropic uncertainty lower bound in the presence of quantum memory   via local operation and classical communication}

\author{F. Adabi}
\affiliation{Department of Physics, University of Kurdistan, P.O.Box 66177-15175, Sanandaj, Iran}
\author{S. Haseli}
\affiliation{Department of Engineering Physics, Kermanshah University of Technology, Kermanshah, Iran}

\author{S. Salimi}
\email{shsalimi@uok.ac.ir}
\affiliation{Department of Physics, University of Kurdistan, P.O.Box 66177-15175, Sanandaj, Iran}

\date{\today}
\begin{abstract}
The uncertainty principle sets lower bound on the uncertainties of  two incompatible observables measured on a particle. The uncertainty lower bound can be reduced by considering a particle as a quantum memory entangled with the measured particle. In this paper, we consider a tripartite scenario in which  a quantum state has been shared between Alice, Bob, and Charlie. The aim of Bob and Charlie is to minimize Charlie's lower bound about Alice's measurement outcomes. To this aim, they concentrate their correlation with Alice in Charlie's side via a cooperative strategy based
on  local operations and classical communication. We obtain lower bound for Charlie's uncertainty about Alice's measurement outcomes after concentrating information and compare it with the  lower bound without concentrating information in some examples.  We also provide a physical interpretation of the entropic uncertainty lower bound based on the dense coding capacity.

\end{abstract}

\maketitle
\section{INTRODUCTION}
The uncertainty principle is one of the most popular and important concepts in quantum theory and  lies at the heart of it \cite{ Heisenberg}. This principle sets limits on the precise prediction of the outcomes of two incompatible quantum measurements on a particle. For example,  the position and momentum of a particle cannot be simultaneously measured with arbitrary high precision in quantum theory. The uncertainty principle can be expressed in different forms. Robertson \cite{Robertson} and Schrodinger \cite{Schrodinger} have shown that for arbitrary pairs of noncommuting observables $Q$ and $R$, the uncertainty principle has the following form
\begin{equation}\label{Roberteq}
\Delta Q \Delta R \geq \frac{1}{2}\vert \langle \left[ Q, R \right] \rangle \vert,
\end{equation}
where $\Delta Q(\Delta R)$ indicates the standard deviation of the associated observable $ Q (R)$
\begin{equation}\label{deviation}
\Delta Q = \sqrt{\langle Q^{2}\rangle -\langle Q \rangle^{2}}, \quad \Delta R = \sqrt{\langle R^{2}\rangle -\langle R \rangle^{2}}.
\end{equation}
An alternative approach to determine the uncertainty relation for any two general observables is based on the entropic measures \cite{WEHNER}. The first version of entropic uncertainty relation was given by Kraus \cite{Kraus} and then proved by Maassen and Uffink \cite{Maassen}, it has the following form
\begin{equation}\label{firstentropy}
H(Q)+H(R)\geq \log_{2} \frac{1}{c},
\end{equation}
where $H(X)=-\sum_{x}p_{x}\log_{2}p_{x}$ is the Shannon entropy of the measured observable $X\in\lbrace Q, R \rbrace$ before the outcome of its measurement is revealed, here $p_{x}$ is the probability of the outcome $x$, $c =\max_{i,j}\vert \langle q_{i} \vert r_{j} \rangle \vert^{2}$ quantifies the ‘complementarity’ between the observables ${Q,R}$ and $\vert q_{i}\rangle$, $\vert r_{j}\rangle$ are the eigenvectors of $Q$ and $R$, respectively. For mixed states, this bound was improved and tightened  \cite{Frank,Berta}  as
\begin{equation}\label{secondentropy}
H(Q)+H(R)\geq \log_{2} \frac{1}{c}+S(\rho),
\end{equation}
where $S(\rho)=-tr(\rho \log_{2} \rho)$ is the von Neumann entropy of $\rho$, with $\rho$ is the density matrix of the measured particle.

    Berta et al. \cite{Berta} consider a situation in which an extra quantum system serving as a quantum memory $B$, has correlation with the measured quantum system $A$. They proved that the uncertainty of Bob, who has access to the quantum memory, about the result of measurements  $Q$ and $R$ on the Alice's system, A,  is bounded by
\begin{equation}\label{berta1}
S(Q\vert B)+S(R\vert B)\geq \log_{2} \frac{1}{c} + S(A \vert B),
\end{equation}
where, $S(A\vert B)=S(\rho^{AB})-S(\rho^{B})$ is the conditional von Neumann entropy,  $S(X \vert B)$ with $X \in\lbrace Q,R \rbrace$ denotes the conditional von Neumann entropies of the post measurement states after measurement on first subsystem in $X$ basis

$$\rho^{XB}=\sum_{i}(\vert x_{i}\rangle\langle x_{i}\vert\otimes\mathbb{I})\rho^{AB}(\vert x_{i}\rangle\langle x_{i}\vert\otimes\mathbb{I}),$$
where $\lbrace \vert x_{i}\rangle \rbrace$'s are the eigenstates of the observable
$X$, and $\mathbb{I}$ is the identity operator. The quantum memory-assisted uncertainty relation in Eq.\ref{berta1} has important and various applications such as witnessing entanglement and cryptographic security \cite{Berta,Prevedel,Li}. Various attempts\cite{Pati,Pramanik,Coles,Liu,Zhang,Pramanik1,Adabi} have been made to generalize and improve  Eq.\ref{berta1}.    We can rewrite Eq.\ref{berta1} as
\begin{equation}\label{berta2}
S(Q\vert B)+S(R\vert B)\geq \log_{2} \frac{1}{c} + S(\rho^{A})-I(A:B),
\end{equation}
where $I(A:B)=S(\rho^{A})-S(A\vert B)$ is the mutual information between $A$ and $B$. It is obvious that the right-hand side (RHS) of Eq.\ref{berta2}, the uncertainty bound (UB), is reduced whenever $A$ and $B$ are correlated; i.e. $I(A:B)>0$. Although  when $A$ and $B$ are separable, then $S(A \vert B)\geq0$ and the minimum of UB is $\log_{2} \frac{1}{c}$, but if $S(A \vert B)<0$ the UB is less then $\log_{2} \frac{1}{c}$. Negativity of $S(A \vert B)$
means inseparability \cite{xu} and shows that there exist entanglement between subsystems $A$ and $B$. Particulary, when they are maximally entangled the UB become trivial.

  For a tripartite scenario in which a quantum state, $\rho_{ABC}$, has been shared between Alice(A), Bob(B) and Charlie(C), due to the monogamy of entanglement, $A$ can not maximally entangled with both $B$ and $C$ , thus Bob and Charlie can not predict the outcomes of  Alice's measurements  exactly \cite{Hu,Renes,C.B}.

    In this manuscript, we consider a situation in which Bob and Charlie can do local operation and  classical communication (LOCC). Bob and Charlie  concentrate their correlation with Alice in one side, for example in Charlie’s side. Thus, the lower bound of Charlie's uncertainty about Alice's measurement outcomes  reduced.  Here we also give a physical interpretation for Berta \emph{et al}'s uncertainty lower bound based through the dense coding capacity\cite{Bennett2,Horodecki,Piani,Winter}.  We can look at bipartite states $\rho_{AB}$ and $\rho_{AC}$ as a resource for dense coding (DC) \cite{Bennett2,Horodecki,Piani,Winter}.

   The paper is organized as follows. In Sec. \ref{2} an operational
meaning of concentrated information(CI) and one-way CI is briefly recalled, we also use this quantity to rewrite the Berta's uncertainty lower bound. In Sec. \ref{example} we give some examples and compare the uncertainty lower bound before and after LOCC.  We present an operational interpretation of Berta's uncertainty lower bound based on quantum dense coding capacity In Sec. \ref{densun}.  The manuscript closes with results and conclusion in Sec. \ref{conclusion}.
\section{CONCENTRATED INFORMATION AND UNCERTAINTY LOWER BOUND}\label{2}
Consider a  tripartite scenario in which, an arbitrary quantum state $\rho_{ABC}$ has been shared between Alice, Bob, and Charlie. They have agreed on two measurements, $Q$ and $R$. Alice measures either $Q$ or $R$ on  $A$ and informs Bob and Charlie of her measurement choice but not the outcomes. Bob and Charlie want to predict the outcomes of the measurements. If communication between them is forbidden then the uncertainties of Bob and Charlie about Alice's measurements outcomes are
\begin{eqnarray}\label{concen1}
S(Q\vert B)+S(R\vert B) &\geq& \log_{2} \frac{1}{c} + S(A \vert B),
\end{eqnarray}
and
\begin{eqnarray}\label{concen}
S(Q\vert C)+S(R\vert C) &\geq& \log_{2} \frac{1}{c} + S(A \vert C),
\end{eqnarray}
respectively. From the strong subadditivity, $$S(\rho^{AB})+S(\rho^{AC})\geq S(\rho^{B})+S(\rho^{C}),$$ one can see that

\begin{equation}\label{subadditivity}
 S(A\vert B)+ S(A \vert C)\geq0
\end{equation}
So, it is impossible for $S(A \vert B)$ and $S(A \vert C)$ to take negative values, simultaneously. Thus, both Bob and Charlie can not predict the outcomes of  Alice's measurements  with uncertainty less than  $\log_{2} \frac{1}{c}$. In particular, when $\rho_{ABC}$ is pure state, $ S(A\vert B)+ S(A \vert C)=0$,  there is a tradeoff relation between their ability to predict the Alice's measurement outcomes.  In other word, the more precisely the Alice's measurement outcomes is predicted by Bob, the less precisely it will be predicted by Charlie and vice versa\cite{Hu}. This is a confirmation of the  monogamy of entanglement, which simply says that the more Bob is entangled with Alice, the less he is with Charlie. Particulary, when the particle $A$ is entangled with both $B$ and $C$,  neither Bob nor Charlie can exactly predict the Alice's measurement outcomes.

In the above scenario, communication between Bob and Charlie was forbidden, but if Bob and Charlie can do LOCC, they  can reduce the uncertainty lower bound  by concentrating their correlation with Alice in one side, for example in  Charlie's side. To do this, Charlie uses an ancillary quantum register $R$, general state is as $\sigma_{ABCR}=\rho_{ABC}\otimes \rho_{R}$, now Bob and Charlie perform an LOCC protocol to maximizes the mutual information between Alice and Charlie. The corresponding maximal mutual information between Alice and Charlie is called concentrated information \cite{Streltsov}
\begin{equation}\label{aaa}
\mathcal{I}(\rho_{ABC})=\max_{\Lambda_{B \leftrightarrow CR}}I^{A:CR}(\sigma^{\prime}),
\end{equation}
where the maximization is taken over all LOCC protocols $\Lambda_{B \leftrightarrow CR}$ and $\sigma^{\prime}=Tr_{B}[\Lambda_{B \leftrightarrow CR}(\sigma)]$. After  concentrating  information, Charlie's uncertainty lower bound is
\begin{eqnarray}\label{finalin}
S(Q\vert C)+S(R\vert C) &\geq & \log_{2} \frac{1}{c}+S(\rho^{A}) - \mathcal{I}(\rho).
\end{eqnarray}
 One can also consider the one-way LOCC protocol where the classical communication is directed from Bob to Charlie, in this case the maximal mutual information is called one-way concentrated information $\mathcal{I}_{\rightarrow}(\rho_{ABC})=\max_{\Lambda_{B \rightarrow CR}}I^{A:CR}(\sigma^{\prime})$. In this case the uncertainty relation is
 \begin{eqnarray}\label{finalin1 way}
S(Q\vert C)+S(R\vert C) &\geq & \log_{2} \frac{1}{c}+S(\rho^{A}) - \mathcal{I}_{\rightarrow}(\rho).
\end{eqnarray}
  Streltsov \emph{et al}.\cite{Streltsov} have obtained  an upper bound for concentrated information as
  \begin{eqnarray}
  \mathcal{I}(\rho)\leq \min\{I^{A:BC}(\rho),S(\rho^{A})+E_{d}^{AB:C}(\rho)\},
  \end{eqnarray}
  therefore,
  \begin{align}\label{finalin2}
S(Q\vert C)+S(R\vert C)&\geq\log_{2} \frac{1}{c}+S(\rho^{A})- \\ \nonumber
& \min\{I^{A:BC}(\rho),S(\rho^{A})+E_{d}^{AB:C}(\rho)\},
\end{align}

where $E_{d}^{AB:C}(\rho)$ is distillable entanglement.

\section{EXAMPLES}\label{example}
In this section we discuss some examples for which the concentrated information has exact form, and  compare the uncertainty lower bounds before and after the concentrating information .
\subsection{\textbf{Example} \textbf{I}}\label{example1}
\begin{figure}[t]
\includegraphics[width=0.50\textwidth]{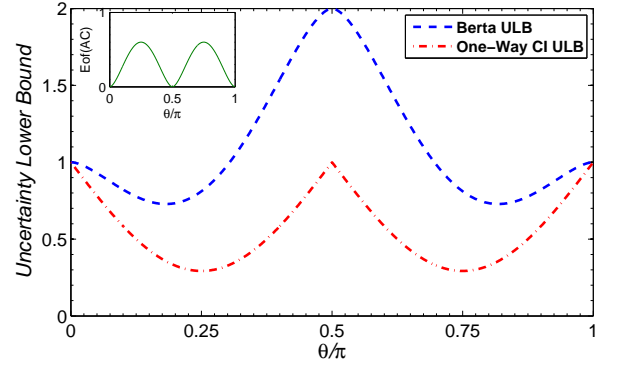}
\caption{\label{figureone}(color online) Lower bounds of the entropic uncertainty relations of the two complementary observables in terms of $\theta /\pi$ , when Alice, Bob and Charli have shared the tripartite state $\vert \psi^{ABC} \rangle =\sin \theta \cos \phi \vert 011\rangle + \sin \theta \sin \phi \vert 101\rangle + \cos \theta \vert 110 \rangle $ with each other. The blue (dashed) line shows the ULB before concentrating information $\log_{2} 1/c + S(A \vert C)$. The red (dot dashed Line) represents ULB after concentrating information. The inset shows the entanglement of formation between $A$ and $C$ . we choose $\phi=\pi / 4$.  }
\end{figure}

We consider the case  where Alice, Bob, and Charlie share a pure state with following form
\begin{equation}\label{psi}
\vert \psi^{ABC} \rangle =\sin \theta \cos \phi \vert 011\rangle + \sin \theta \sin \phi \vert 101\rangle + \cos \theta \vert 110 \rangle .
\end{equation}
In this case, the one-way CI has been obtained exactly by \cite{Streltsov}
\begin{equation}\label{one-way con}
\mathcal{I}_{\rightarrow}(\rho_{ABC})=S(\rho_{A})+E_{a}(\rho_{AC}),
\end{equation}
where $\rho_{ABC}=\vert \psi^{ABC} \rangle\langle \psi^{ABC} \vert$, $E_{a}$ is the entanglement of assistance which is given by\cite{DiVincenzo}
\begin{equation}
E_{a}(\rho_{AC})=\max \sum_{i} p_{i} E_{d}(\vert \psi_{i}^{AC} \rangle),
\end{equation}
The maximum takes over all possible pure-state decompositions of $\rho_{AC} = Tr_{B}(\rho_{ABC})=\sum_{i} p_{i} \vert \psi_{i} \rangle_{AC}\langle \psi_{i} \vert$, for a pure state $\vert \psi_{i}^{AC} \rangle$  the distillable entanglement is equal to the von Neumann entropy of the reduced state i.e. $ E_{d}(\vert \psi_{i}^{AC} \rangle)=S(\rho_{i}^{A})$. In Fig.\ref{figureone} we plot the uncertainty lower bound (ULB) of enropic uncertainty relations in Eqs.\ref{concen} and \ref{finalin1 way} in terms of $\theta / \pi$ with $\phi= \pi /4$. As expected, after concentrating information the uncertainty lower bound   is reduced. As can be seen in Fig.\ref{figureone}, when the entanglement between parts $A$ and $C$ increases the uncertainty lower bound decreases.
\begin{figure}[t]
\includegraphics[width=0.45\textwidth]{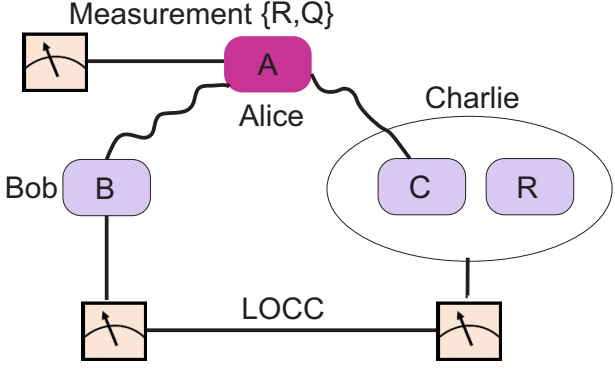}
\caption{\label{figasli}Schematic representation of our setting where  the quantum memory $C$ (Charlie's system) has access
to a quantum register $R$. Bob and Charlie perform local operations
and classical communication (cooperative strategy) to maximize mutual information between Alice and Charlie. Alice does measurement on her system  }
\end{figure}
\subsection{\textbf{Example} \textbf{II}}\label{example2}
\begin{figure}[t]
\includegraphics[width=0.50\textwidth]{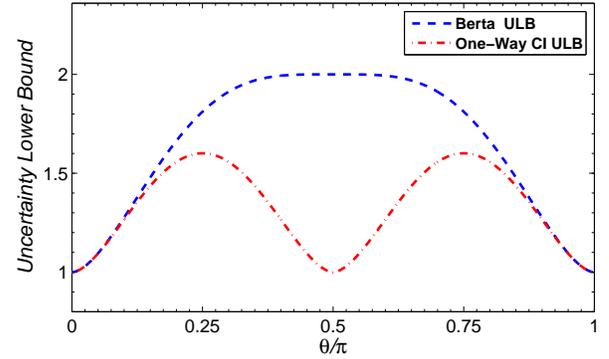}
\caption{\label{figuretwo}(color online) Lower bounds of the entropic uncertainty relations of the two complementary observables in terms of $\theta /\pi$ , when Alice, Bob and Charli have shared the tripartite state $(\sin^{2}\theta \vert \phi \rangle_{AB}\langle \phi \vert + \cos^{2} \theta \vert) \otimes (p \vert 0 \rangle_{C}\langle 0 \vert + (1-p) \vert 1 \rangle_{C}\langle 1 \vert) $ with each other. The blue-dashed line shows the ULB before concentrating information, $\log_{2} 1/c + S(A \vert C)$. The red-dot dashed line represents ULB after concentrating information. we choose $p=1/3$  }
\end{figure}

As a second example, we consider a tripartite state of the form $\rho_{AB}\otimes \rho_{C}$. In these case there is no correlation between Alice's and Charlie's part. Charlie can learn about Alice by asking Bob. It can be said explicitly that Charlie can improve its uncertainty about Alice's measurement outcomes with the help of Bob. In this case, the CI and the one-way CI is equal, and are given by \cite{Streltsov}
\begin{equation}
\mathcal{I}(\rho_{AB}\otimes \rho_{C})=\mathcal{I}_{\rightarrow}(\rho_{AB}\otimes \rho_{C})=I(A:B)-D(A\vert B),
\end{equation}
where $D(A\vert B)$ is quantum discord\cite{Zurek}. We study a case in which Alice, Bob and Charlie share a tripartite state of the following form
\begin{eqnarray}\label{mesal2}
\rho_{ABC}&=&\rho_{AB}\otimes \rho_{C},\\ \nonumber
\rho_{AB}&=& \sin^{2}\theta \vert \phi \rangle_{AB}\langle \phi \vert + \cos^{2} \theta \vert 1,1\rangle_{AB}\langle 1,1\vert , \\ \nonumber
\rho_{C}&=& p \vert 0 \rangle_{C}\langle 0 \vert + (1-p) \vert 1 \rangle_{C}\langle 1 \vert ,
\end{eqnarray}
where $\vert \phi \rangle = 1/\sqrt{2} (\vert 0,1 \rangle + \vert 1,0 \rangle) $, with $0 \leq p \leq 1$. As can be seen from Fig.\ref{figuretwo} the uncertainty lower bound after concentrating information is reduced, although there is not any correlation between parts $A$ and $C$ initially.
\subsection{\textbf{Example} \textbf{III}}\label{example4}
\begin{figure}[t]
\includegraphics[width=0.50\textwidth]{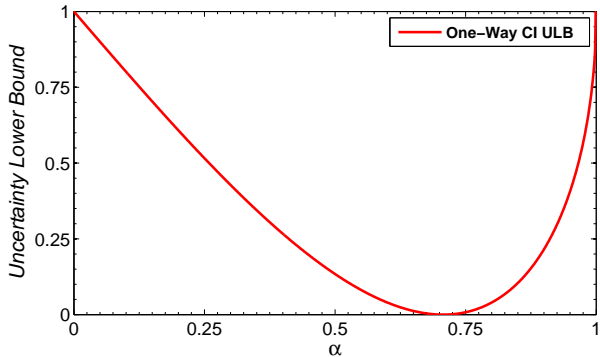}
\caption{\label{figuretree}(color online) Lower bounds of the entropic uncertainty relations of the two complementary observables in terms of $\alpha$ , when Alice, Bob and Charli have shared the (GHZ) tripartite pure state $\vert GHZ \rangle=\sqrt{\alpha^{2}}\vert 0,0,0 \rangle + \sqrt{1-\alpha^{2}}\vert 1,1,1 \rangle$ .The ULB before concentrating information is $\log_{2} 1/c + S(A \vert C)=1$. The red -solid represents ULB after concentrating information.}
\end{figure}

As a last example we consider a case where Alice, Bob and Charlie have shared the GHZ tripartite pure state $\vert GHZ \rangle=\sqrt{\alpha^{2}}\vert 0,0,0 \rangle + \sqrt{1-\alpha^{2}}\vert 1,1,1 \rangle$, with $ 0\leq\alpha\leq1$. This state is pure, so the one-way CI can be exactly obtained according to Eq. \ref{one-way con}. Here it is easy to show that for all $\alpha$, the uncertainty lower bound  before concentrating information is equal to one. However, after one-way concentrating information the ULB is reduced. In Fig.\ref{figuretree} the plot shows the ULB after  concentrating information. As can be seen from Fig.\ref{figuretree} the uncertainty lower bound after  one-way concentrating information is smaller than one and for $\alpha=\frac{1}{\sqrt{2}}$ the UB is equal to zero.

\section{DENSE CODING CAPACITY AND UNCERTAINTY LOWER BOUND }\label{densun}
Quantum dense coding  was first proposed by Bennett and Wiesner \cite{Bennett2}. In the dense coding protocol, entangled states are used to transmit classical information through a quantum channel. Actually, DC is a task that uses pre-established entanglement between sender and receiver to send classical messages more efficiently. Let Alice and Bob, share the pre-established entangled quantum state $\rho_{AB}$ in composite Hilbert space $H_{A} \otimes H_{B}$, where $H_{A}(H_{B})$ is the Hilbert space of Alice(Bob). Alice wants to use this quantum state as a channel for sending classical information to Bob. In general (DC) \cite{Horodecki,Piani,Winter}, Alice encodes
her classical message by means of general quantum operations
\begin{equation}\label{encode}
\rho_{\tilde{A}B}=(\Lambda_{A} \otimes \mathbb{I}) \rho_{AB},
\end{equation}
the quantum operation changes the dimension of $A$ from $d_A$ to $d_{\tilde{A}}$ ( $d_{\tilde{A}}$ is the dimension of the subsystem sent to $B$) If the encoding procedure is applied for single copies of $\rho_{AB}$, the (DC) capacity is given by \cite{Cavalcanti}
\begin{equation}\label{DCcapacity}
C_{DC}(A\rangle B)=\log_{2}d_{\tilde{A}} + \max_{\lbrace \Lambda_{A}\rbrace} I(\tilde{A}\rangle B),
\end{equation}
where $I(\tilde{A}\rangle B)$ is the coherent information
of $\rho_{\tilde{A}B}$ and the optimization is over all quantum operations $\Lambda_{A}$ with output dimension $d_{\tilde{A}}$.
Since dense coding help us to present an operational interpretations for quantum discord (QD)\cite{Zurek}, we do an overview of quantum discord. For bipartite state $\rho_{AB}$, the classical correlation is given by $J(B\vert A)=S(\rho_{B})-\min_{\lbrace E_{k}^{A}\rbrace}S(\rho^{B\vert \lbrace E_{k}^{A}\rbrace})$(Provided that the (POVM) measurement $\lbrace E_{k}^{A}\rbrace$ is performed on the first part), here $\rho^{B\vert \lbrace E_{k}^{A}\rbrace}=tr_{A}(E_{k}^{A}\rho_{AB})$ is postmeasurement state with probability for obtaining the outcome $k$ as $p_{k}=tr(E_{k}^{A}\rho_{AB})$. Thus quantum discord is defined as
\begin{equation}\label{discord}
D(B\vert A)=I(A:B)-J(B\vert A),
\end{equation}
where $I(A:B)=S(A)+S(B)-S(AB)$ is quantum mutual information.
As already mentioned, in comparison with first entropic uncertainty relation in Eq.\ref{firstentropy} The uncertainty relation in Eq.\ref{berta1} has additional term $S(A \vert B)$. Here we provide a physical interpretation for this additional term, it is expressed based on DC capacity. In Ref.\cite{fanchini} Fanchini et al. show that for tripartite scenario with a pure quantum state $\rho_{ABC}$, conditional von Neumann entropy $S(A \vert B)$ can be expressed as
\begin{equation}\label{compett}
S(A \vert B)=D(C\vert A)-D(B\vert A)
\end{equation}
this relation has also been shown in Ref.\cite{Hu}. It can be easily proved that the following equation is established between QD and DC capacity\cite{Modi,Cavalcanti}
\begin{equation}\label{final}
D(C\vert A)-D(B\vert A)=C_{DC}(A\rangle C) - C_{DC}(A\rangle B).
\end{equation}
From Eq.\ref{compett} and Eq.\ref{final}, it can be concluded that the second term in the RHS of Eq.\ref{berta1} has the following form
\begin{equation}\label{ub}
S(A \vert B)=C_{DC}(A\rangle C) - C_{DC}(A\rangle B).
\end{equation}
It indicate that, when the DC capacity from Alice (sender) to Bob(receiver) is greater than the DC capacity from Alice (sender) to Charlie(receiver) then Bob's  uncertainty lower bound is less than Charlie's uncertainty lower bound  and vice versa.
\section{CONCLUSION}\label{conclusion}
It was shown that  the entropic uncertainty lower bound can be reduced by considering a particle as a quantum memory entangled with the primary particle.  For a tripartite scenario in which a quantum state, $\rho_{ABC}$, has been shared between Alice(A), Bob(B) and Charlie(C), due to the monogamy of entanglement, $A$ can not maximally entangled with both $B$ and $C$ , thus Bob and Charlie can not predict the outcomes of  Alice's measurements  exactly. Here,
we have shown that  Bob and Charlie can reduce the uncertainty lower bound by concentration information via a cooperative strategy based on  local operations and classical communication. We have compared the uncertainty lower bound before and after concentrating information for some examples.  We also  have provided a physical interpretation of Berta's uncertainty lower bound in the scenario with three players. The second term of Berta's uncertainty lower bound in Eq.\ref{berta1}  can be expressed as difference between DC capacity from Alice (sender) to Charlie(receiver)  and DC capacity from Alice (sender) to Bob(receiver), such that when the DC capacity from Alice (sender)
to Bob(receiver) is greater than the DC capacity from Alice
(sender) to Charlie(receiver) then entropic uncertainty lower
bound of Bob about Alice's measurement outcome is reduced and vice versa.

\end{document}